\def\BML{\begin{mathletters}}
\def\EML{\end{mathletters}}
\def\BE{\begin{equation}}
\def\EE{\end{equation}}
\def\BEA{\begin{eqnarray}}
\def\EEA{\end{eqnarray}}
\def\NN{\nonumber}
\def\L{\langle}
\def\R{\rangle}

\documentstyle[twocolumn,pre,aps]{revtex}
\input epsf.tex
\begin{document}
\twocolumn[
\hsize\textwidth\columnwidth\hsize\csname@twocolumnfalse\endcsname

\title{Thermal phase diagrams of columnar liquid crystals}
\author{G. Lamoureux, A. Caill\'e and D. S\'en\'echal}
\address{D\'epartement de physique and Centre de recherche en physique du solide,}
\address{Universit\'e de Sherbrooke, Sherbrooke, Qu\'ebec, Canada J1K 2R1.}
\maketitle
\begin{abstract}%
In order to understand the possible sequence of transitions from the
disordered columnar phase to the helical phase in hexa(hexylthio)triphenylene
(HHTT), we study a three-dimensional planar model with octupolar interactions
inscribed on a triangular lattice of columns. We obtain thermal phase diagrams
using a mean-field approximation and Monte Carlo simulations. These two
approaches give similar results, namely, in the quasi one-dimensional regime,
as the temperature is lowered, the columns order with a linear polarization,
whereas helical phases develop at lower temperatures. The helicity patterns of
the helical phases are determined by the exact nature of the frustration in
the system, itself related to the octupolar nature of the molecules. 
\end{abstract}
\pacs{}
]

\narrowtext
\tightenlines

\section{Introduction}

The study of phase transitions in columnar liquid
crystals\cite{Chandrasekhar,deGennes} presents a fundamental interest in that
these materials combine, aside from the vast phenomenology of soft matter,
many features at the origin of important phenomena of the solid state, namely
a relatively strong elastic anisotropy in the direction of the columns, a
geometrical frustration of the intermolecular interaction coming from the
triangular nature of the lattice of columns, and finally discoid molecules
with nontrivial point-group symmetry. The present study is based on
hexa(hexylthio)triphenylene (HHTT), a compound formed of a rigid core of
aromatic cycles, giving a discoid shape to the molecule, and of six flexible
hydrocarbon chains, responsible for its characteristic thermotropic character.
HHTT is the only molecule from the triphenylene derivatives to show two
distinct columnar phases. Indeed, as the temperature decreases, the sequence
of phases is the following: $I$, an isotropic liquid; $D_{hd}$, a disordered
columnar phase; $H$, a helically ordered columnar phase and finally $K$, a
monoclinic crystal. These phases were identified by X-rays measurements on
powders\cite{Gramsbergen86} and freely suspended
strands\cite{Fontes88,Fontes89,Idziak92} of HHTT.

These X-rays results are best interpreted by concluding that the $D_{hd}$
phase of HHTT ($70^\circ$C$<T<93^\circ$C) has long-range positional order in
the plane perpendicular to the columns: columns are located on a triangular
lattice. Within a column, short range (liquid) positional order is realized.
The columns slide freely one against the other. The $H$ phase
($62^\circ$C$<T<70^\circ$C) has in-column long-range positional and
orientational helical order, or quasi long-range order, as proposed in
\onlinecite{Caille96}. In this last phase, two neighboring molecules in a
single column are separated on the average by a distance $d_\|=3.6${\AA} and
rotated from each other by an angle $\alpha\approx 45^\circ$, constant on the
whole temperature interval of the phase. In order to reduce the frustration
associated with the triangular geometry of the lattice, the lattice
reorganizes itself in a superlattice  $\sqrt{3}\times\sqrt{3}R30^\circ$: one
third of the columns have a vertical offset of half the inter-molecular
distance ($d_\|/2$). The displaced columns have an opposite helicity:
$\alpha\approx -45^\circ$, instead of $+45^\circ$ for the undisplaced columns.
If it were not for the very high value of the mean square displacement of the
molecules in the direction of the columns, the $H$ phase would seem
very similar to a crystalline phase. In that sense, the exact mechanism of the
$D_{hd}\leftrightarrow H$ transition is still an open question. In an effort
towards elucitating the mechanism, we study the orientational ordering
assuming from the start that the HHTT molecules already occupy well defined
positions.

We study the thermal phase diagram of a model Hamiltonian by means of a Landau
free energy functional in a mean-field approximation and of Monte Carlo
simulations on finite size lattice. Previous work has been done on the ground
state of a related model\cite{Hebert96} and on thermal phase diagrams for a
two-dimensional model of uniform columns\cite{Hebert95,HebertPlumer96}. Our
analysis confirms that, as previously seen at $T=0$, the octupolar $G$
coupling\cite{Plumer93} is determinant for obtaining the helicity
configuration of the columns at any temperature. It also shows that a
diversity of phases survives at $T\ne 0$. For weak transverse couplings, the
model produces the expected low-temperature helical phases, but also suggests
that some linearly polarized phases could exist at higher temperatures.

In the following section, the model Hamiltonian is presented with an emphasis
on the inter-columnar intermolecular interactions. In Sec. III, using
mean-field approaches, the locus of points for the second order phase
transitions from the disordered phase to an ordred phase is obtained. It is
followed by a determination of the thermal phase diagrams and characterisation
of the different phases in terms of the helicity pattern and relative phases.
In Sec. IV, the thermal phase diagrams are obtained using Monte Carlo
simulations on finite size systems in conjunction with the spiraling
algorithm. Finally, Sec. V discusses the results and arrives at general
conclusions.

\section{Model Hamiltonian}

As indicated it the introduction, the main purpose of our calculation is to
elucidate the role played by the angular degrees of freedom. Even though this
may be questionable formally, we will freeze the positional degrees of freedom
to render the problem tractable. Despite the fact that the exact positional
order in the columns is not known, we then assume that the molecules lie on a
three-dimensional triangular lattice of ordered columns (with one third of the
columns being displaced). This simplification is certainly valid in the $H$
phase, even if only quasi long-range positional order existed since order
would then be maintained over many inter-molecular distances. Accordingly, we
will use $(i,j)$ site indices to identify the unit cell of the three columns
of the two-dimensional $\sqrt{3}\times\sqrt{3}R30^\circ$ superlattice of
columns and $\mu$ to identify the column: $\mu=1$ and 2 label the undisplaced
columns and $\mu=3$ labels the columns offset by $d_\|/2$. Finally, an index
$k$ identifies the sites a molecule occupies in the $(\mu,i,j)$ column. We
formally write $(i,j,k)$ as $m$, an index labelling a plaquette of three
molecules.

\begin{figure}
\vglue 0.4cm\epsfxsize 8cm\centerline{\epsfbox{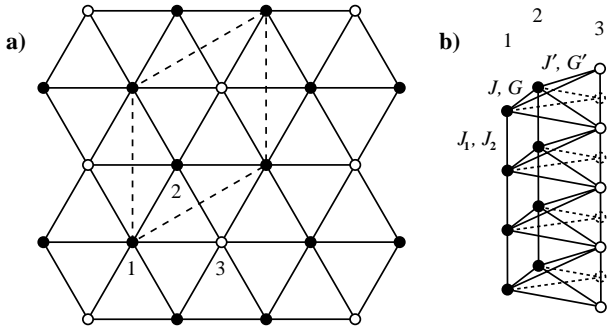}}\vglue 0.4cm
\caption{\baselineskip=12pt
a) Two-dimensional triangular lattice of columns. Empty dots represent displaced columns ($\mu=3$) and filled dots, undisplaced columns ($\mu=1$ and 2). Dashed lines represent the elementary cell of the super-lattice. b) Three-dimensional illustration of the couplings for three columns. Dotted lines represent fictive sites and interactions for an undisplaced column 3.
}\label{reseau}
\end{figure}

Within the framework of this plastic state model, each molecule $(\mu,m)$ has
a well-defined position ${\bf r}_{\mu m}$ and an orientation labelled
$\theta_{\mu m}$, admitting from the beginning that its plane is perpendicular
to the direction of the columns. An orientational disorder may mimick an
effective $D_{hd}$ phase, which however would possess orientational and
positional disorders along the columns. We did
not explicitly consider the shape and flexibility of the tails, which would
vary with the temperature. Recall, however, that it has been
suggested\cite{Fontes88,Fontes89} that the stiffening of the tails may be
responsible for the $D_{hd}\rightarrow H$ transition. The above
considerations are implicitly integrated out as weakly temperature-dependent
renormalization effects of the inter-columnar interactions, allowing us to
use, to a good approximation, an effective model with athermal values of the
interaction parameters. The resulting thermal phase diagrams will represent a
somewhat distorted version of the true temperature dependence.

The last thing to consider in the model is that the molecules are not exactly
disk-like: they have a chirality associated with the alternate arrangement of
the aliphatic tails. Indeed, conformational analysis on compounds similar to
HHTT\cite{Pesquer80,Cotrait82} shows that this ``propeller blade''
configuration is the ground state of a single molecule and of two stacked
molecules, one on top of the other.

\subsection{Intra-columnar interactions}

In their ground state, two stacked HHTT molecules minimize their conformation
energy by allowing an angular shift $\alpha$ between the two
molecules\cite{Pesquer80,Cotrait82}. An intrinsic chiral model represented by
the Hamiltonian
\BE
-J\cos 3(\theta_{k+1}-\theta_k-\alpha),
\EE
would be appropriate to represent this situation since the HHTT molecules have
$D_3$ symmetry. However there is no a priori selection between the right-handed
or left-handed chirality. In order to allow for the two possible chiralities
of each column (the sign of $\alpha$), we preferably use a
next-nearest-neighbor model (see \onlinecite{Selke92,RednerStanley77}), with
competing interactions of the form:
\BE
\label{J1J2coupling}
-J_1\cos 3(\theta_{k+1}-\theta_k)
-J_2\cos 3(\theta_{k+2}-\theta_k),
\EE
with $J_1>0$ and $J_2<0$. In the ground state, for $4|J_2|\ge|J_1|$, the
molecules adopt a helical configuration with pitch $\alpha$, given by
$\cos3\alpha=-J_1/4J_2$. Otherwise, the intrinsinc helicity of the columns is
zero. The $J_1/J_2$ ratio determines the magnitude of $\alpha$, but allows
opposite helicities for different columns or even helicity reversals within a
column, separating helicity domains.

For an isolated molecule of HHTT at finite temperature, it is unclear that
chirality is a well defined property: conformational analysis\cite{Pesquer80}
shows that the energy barrier between opposite chirality configurations is
comparable to the thermal energy in the $H$ and $D_{hd}$ phases. Thus, instead
of using an additional Ising variable on each site to represent the chirality
of the molecule, we represent the chirality as the result of a collective
behavior in the effective model (\ref{J1J2coupling}), submitted to
inter-columnar interactions: every molecule in a particular helicity domain
has the same chirality, related to the sign of the helicity. The
inter-columnar interactions are essential to stabilize helicity domains at
finite temperature.

\subsection{Inter-columnar interactions}

Given the approximation that each molecule is fully described by its
orientation $\theta_{\mu m}$, we may write its mass density as a multipole
expansion\cite{Plumer93,Choi89}. Because of the $D_3$ point symmetry of the
molecule, the first nonzero moment is the octupolar moment, which may be
represented by $Q_{klm}$, a rank-three tensor ($k,l,m=x,y$). The only
interactions that are bilinear in $Q$ as well as invariant with respect to the
symmetries of the hexagonal lattice have a $\cos 3(\theta-\theta')$ or $\cos
3(\theta+\theta')$ form\cite{Plumer93}. The inter-columnar interaction is
then approximated to be
\BE
\label{JGcoupling}
-J\cos 3(\theta_{\mu ij}-\theta_{\mu'i'j'})
-G\cos 3(\theta_{\mu ij}+\theta_{\mu'i'j'}).
\EE
The first term is invariant under continuous rotations and would be present
even if the molecules had lower symmetry multipole moments. However, the
second term is specific to the octupolar character of the molecules and has
only discrete rotational symmetries.

To extract the $D_3$ symmetry of the molecules, we replace the real
orientations $\theta_{\mu m}$ by angular variables $\phi_{\mu m}=3\theta_{\mu
m}$. The complete hamiltonian of the system then reads:
\BEA
\label{hamilCos}
H = -\sum_{\mu\nu} \sum_{mn}
[&&J_{\mu m,\nu n}\cos(\phi_{\mu m}-\phi_{\nu n}) \NN\\
&&+G_{\mu m,\nu n}\cos(\phi_{\mu m}+\phi_{\nu n})].
\EEA
$J_{\mu m,\nu n}$ contains the intra-columnar interactions: each site is
coupled to its first and second intra-column nearest neighbors by $J_1$ and
$J_2$ as in Eq.~(\ref{J1J2coupling}). The inter-columnar couplings are
embedded both in $J_{\mu m,\nu n}$ and $G_{\mu m,\nu n}$. Each $\mu=1$ site
interacts with three $\mu=2$ and six $\mu=3$ neighbors, three upwards and three
downwards. The inter-columnar couplings have different values: $J$ and $G$ for
in-plane molecules (1-2 bonds) and $J'$ and $G'$ for out-of-plane molecules
(1-3 and 2-3 bonds) (see Fig.~\ref{reseau} b). Nevertheless, it is physically
justified\cite{Hebert95} to suppose that $J'\approx J$ and $G'\approx G$, and
for simplicity, we assume $J'=J$ and $G'=G$. $J_1$ is positive and taken to be
unity (it sets the energy scale).

With the notations
\BE
c_{\mu m} = \cos(\phi_{\mu m})
\quad\text{and}\quad
s_{\mu m} = \sin(\phi_{\mu m}),
\EE
we may re-write the hamiltonian (\ref{hamilCos}) as
\BE
\label{hamilCS}
H = -\sum_{\mu\nu} \sum_{mn}
[J^c_{\mu m,\nu n} c_{\mu m} c_{\nu n}
+J^s_{\mu m,\nu n} s_{\mu m} s_{\nu n}],
\EE
where $J^c_{\mu m,\nu n} = J_{\mu m,\nu n} + G_{\mu m,\nu n}$ and $J^s_{\mu
m,\nu n} = J_{\mu m,\nu n} - G_{\mu m,\nu n}$. The reader should note that the
$G\leftrightarrow-G$ transformation interchanges the $c$ and $s$ variables and
is equivalent to a rotation $\phi_{\mu m}\leftrightarrow\phi_{\mu
m}+\case{\pi}{2}$ of the molecules.

\section{Mean-field calculation}

We use a six-component variable
$S_{im}=(c_{1m},c_{2m},c_{3m},s_{1m},s_{2m},s_{3m})$. In a Fourier
representation, the mean-field is
\BE
\label{champMoyen}
h_i({\bf q}) = \sum_j J_{ij}({\bf q}) \L S_j({\bf q})\R,
\EE
where $J_{ij}({\bf q})$ is a $6\times 6$ block diagonal matrix constructed from
the $J^c$ and $J^s$ couplings of Eq. (\ref{hamilCS}):
\BE
J({\bf q}) = \pmatrix{
J^c({\bf q}) & 0 \cr
0 & J^s({\bf q}) \cr}.
\EE
Because there are three columns in the unit cell, the mean-field transverse
components of ${\bf q}$ are zero, as we have verified through a detailed
calculation. From now on, without any ambiguity, we replace ${\bf q}$ by $q$,
its $z$ component. We also take the intra-column distance between two
molecules $d_\|=1$. The $J^c$ and $J^s$ matrices are then
\BE
J^{c,s}(q) = \pmatrix{
J_\|(q)         & J^{c,s}_{12}    & J^{c,s}_{31}(q) \cr
J^{c,s}_{12}    & J_\|(q)         & J^{c,s}_{23}(q) \cr
J^{c,s}_{31}(q) & J^{c,s}_{23}(q) & J_\|(q)         \cr},
\EE
with $J_\|(q) = \cos q+J_2\cos 2q$, $J^c_{12} = \case3/2(J+G)$, $J^c_{23}(q) =
J^c_{31}(q) = 3(J+G)\cos\case1/2 q$, $J^s_{12} = \case3/2(J-G)$ and
$J^s_{23}(q) = J^s_{31}(q) = 3(J-G)\cos\case1/2 q$. The displacement of the
$\mu=3$ columns changes the coordination number from 6 to 3 and adds a
$\cos\case1/2 q$ factor.

\subsection{Second-order phase transition temperature $T_c$}

The order-disorder continuous phase transitions are related to the divergence
of the ``paramagnetic'' susceptibility $\chi$, which in turn is related to the
single-site susceptibility $\chi_0=1/2T$, with $k_B=1$, by the standard RPA
relation:
\BE
\chi(q) = \chi_0 [\openone - \chi_0 J(q)]^{-1}.
\EE
The $6\times 6$ matrix between brackets is non-invertible when at least one
of its eigenvalues is zero. As the temperature is lowered, the transition
occurs for some $q_c$ maximizing one of the six eigenvalues of $J(q)$. The
corresponding eigenvector identifies the configuration involved in the
transition. The eigenvalue itself is twice the critical temperature $T_c$.

$J(q)$ is block-diagonal, and the 6th-order caracteristic equation reduces to
the two cubic equations given by
\BE
\det(J^{c,s}(q)-j^{c,s}\openone) = 0.
\EE
The eigenvalues are
\BML
\BE
j^{c,s}_1(q) = J_\|(q) - J^{c,s}_{12},
\EE
\BE
j^{c,s}_2(q) = J_\|(q) + \case1/2\Bigl[ J^{c,s}_{12}
- \sqrt{ {J^{c,s}_{12}}^2 + 8{J^{c,s}_{23}(q)}^2 } \Bigr],
\EE
\BE
j^{c,s}_3(q) = J_\|(q) + \case1/2\Bigl[ J^{c,s}_{12}
+ \sqrt{ {J^{c,s}_{12}}^2 + 8{J^{c,s}_{23}(q)}^2 } \Bigr].
\EE
\EML
The eigenvectors are of the form
\BML
\BE
v^{c,s}_1(q) = ( 1, -1, 0 ),
\EE
\BE
v^{c,s}_2(q) = ( 
j^{c,s}_2 - J_\|(q),\,\,
j^{c,s}_2 - J_\|(q),\,\, 2 J^{c,s}_{23}(q) ),
\EE
\BE
v^{c,s}_3(q) = (
j^{c,s}_3 - J_\|(q),\,\,
j^{c,s}_3 - J_\|(q),\,\, 2 J^{c,s}_{23}(q) ).
\EE
\EML
For each parameter set $(J,G,J_2)$, we numerically find which of the six
eigenvalues is maximal and the corresponding $q_c$. It turns out that the only
two eigenvalues to be maximum are $j^c_3$ and $j^s_3$. The critical
temperature is thus the maximum of the following two temperatures:
\BEA
\label{relationTc}
T_c^{c,s} =&& \max(\case1/2 j^{c,s}_3) \NN\\
=&& \case1/2(\cos q_c+J_2\cos 2q_c)+\case3/8(J\pm G) \\
&& +\case3/8|J\pm G|\sqrt{1+32\cos^2 \case1/2 q_c}. \NN
\EEA
If $T_c^c > T_c^s$ (resp. $T_c^c < T_c^s$), only the cosine (resp. sine)
components are involved in the transition. The boundary between the cosine and
sine transitions is defined by $T_c^c = T_c^s$. Noticing the $(a,a,b)$
structure of the $v^c_3$ and $v^s_3$ vectors, we conclude that columns 1 and 2
play similar roles, while column 3 has a distinct behavior.

\begin{figure}
\vglue 0.4cm\epsfxsize 8cm\centerline{\epsfbox{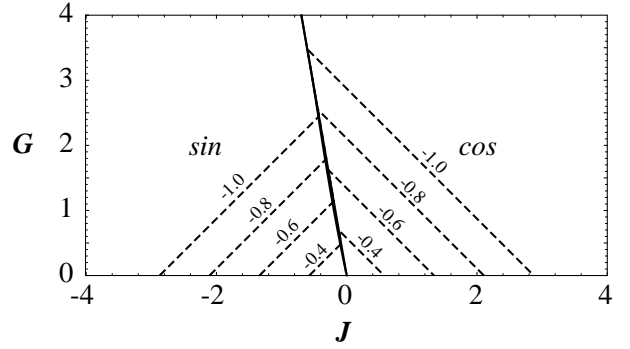}}\vglue 0.4cm
\caption{\baselineskip=12pt
Phase diagrams at the critical temperature for $J_2=-1.0$, $-0.8$, $-0.6$ and
$-0.4$. The full lines, all superimposed with the
precision used, are the boundaries between {\it cos} and {\it sin} phases and
the dashed lines, the boundaries between $q_c=0$ and $q_c\ne 0$ phases. For
$G<0$, the boundaries are a mirror image of the above with {\it cos} and {\it
sin} phases being interverted. }\label{diagTc}
\end{figure}

In the $J$-$G$ plane, and for a particular value of $J_2$, we identify four
regions corresponding roughly to the four quadrants, two of them being shown
in Fig.~\ref{diagTc}. The line $G=0$, where the cosine and sine components are
equivalent, is an obvious boundary. The curve
\BE
|G|=-g(J,J_2)J,
\EE
on which $T_c^c = T_c^s$, follows from a mechanism similar to the spin-flop
mechanism of magnetism. For $|J|$ and $|G|$ sufficiently high relative to
$|J_2|$, $q_c$ vanishes, that is, the transverse couplings have ``untwisted''
the columns. The equation $T_c^c(q_c=0) = T_c^s(q_c=0)$ then gives a relation
between $J$ and $G$:
\BE
\label{relationJG}
(J+G)+|J+G|\sqrt{33} = (J-G)+|J-G|\sqrt{33}.
\EE
The solutions are $G=0$ and $|G|=-\sqrt{33}J$. Thus, in this limit of $q_c=0$,
$g=\sqrt{33} \approx 5.74$, independently of $J$ or $J_2$. At $q_c=0$, the
columns behave like a single vector flipping under the anisotropic effect of
$G$. For smaller values of $|J|$ and $|G|$, $q_c\ne 0$ and $g$ decreases, the
same mechanism however remains.

In the ``cosine'' regions, the helicity at the transition is a function
$q_c=q_c(J+G,J_2)$, because $J+G$ is the only combination of $J$ and $G$
appearing in $j^c_3$. In the ``sine'' regions, we have $q_c=q_c(J-G,J_2)$ for
the same reason. For $J_2>-\case1/4$, $q_c=0$, that is, each column stabilizes
a ``ferromagnetic'' order\cite{Selke92,RednerStanley77}. For $J_2<-\case1/4$,
there is a region of the $J$-$G$ plane where $q_c\ne 0$, but $q_c=0$ for $|J|$
and $|G|$ high enough. The boundary is determined by the competition between
$|J_2|$ and the transverse couplings, respectively inducing a modulation in the
columns, and favoring $q_c=0$.

\subsection{Thermal phase diagrams}
In order to investigate finite-temperature effects near the order-disorder
transition, to better specify the nature of the phases, and to rule out the
possibility of higher temperature first-order transitions, we set up a Landau
theory from the microscopic model. We follow in essence the method proposed by
Bak and von Boehm \cite{Bak80}.

\subsubsection{Free energy functional expansion}
For commodity, we divide the Landau free energy functional $F$ into two parts:
$F_T$ and $F_J$. In reciprocal space, to fourth order, we find that
\BEA
F_T = &&T \sum_\mu \Bigl\{
\sum_{q_1} \Bigl[
c_\mu(q_1) c_\mu(-q_1)
+ s_\mu(q_1) s_\mu(-q_1) \Bigr] \NN\\
&&+ \case1/4 \sum_{q_1 q_2 q_3} \Bigl[
c_\mu(q_1) c_\mu(q_2) c_\mu(q_3) c_\mu(-q_1-q_2-q_3) \NN\\
&&
+ 2 c_\mu(q_1) c_\mu(q_2) s_\mu(q_3) s_\mu(-q_1-q_2-q_3) \NN\\
&&
+ s_\mu(q_1) s_\mu(q_2) s_\mu(q_3) s_\mu(-q_1-q_2-q_3) \Bigr]
\Bigr\}
\EEA
with $T=1/\beta$. $c_\mu(\tilde{q})$ and $s_\mu(\tilde{q})$ are the Fourier
transforms of the mean values $\L c_{\mu m}\R$ and $\L s_{\mu m}\R$. This
limited power expansion is numerically close to (less than 1\% difference) the
exact value up to $\sqrt{\L c_{\mu m}\R^2 + \L s_{\mu m}\R^2} \approx 0.5$. A
6th-order development is 1\% accurate up to $\sqrt{\L c_{\mu m}\R^2 + \L
s_{\mu m}\R^2} \approx 0.65$. For every mean value under this limit of
validity, a negligible number of fictitious spins with modulus higher than 1
contributes to the statistics. We also find that
\BEA
F_J = - \sum_{\mu\nu} \sum_{q_1} \Bigl[
&&J_{\mu\nu}^c(q_1) c_\mu(q_1) c_\nu(-q_1) \NN\\
&&+ J_{\mu\nu}^s(q_1) s_\mu(q_1) s_\nu(-q_1)
\Bigr].
\EEA
These expressions represent the free energy functional for a group of three
columns. The $q$ summations have been truncated of all the umklapp terms.
These umklapp terms would have pinned the modulation to commensurate values.
By ignoring them, we allow incommensurate phases to occupy the entire
parameter space, leaving a space of measure zero to commensurate phases.
In real systems a devil's staircase\cite{Bak80} is expected
instead of the continuous $q$ profile.

\subsubsection{Order parameters}
We then assume that, near the transition, $c_\mu(\tilde{q}) = 0$ and
$s_\mu(\tilde{q}) = 0$ ($\forall\mu$) except for $\tilde{q} = \pm q$. In other
words, we concentrate on the first harmonic to appear in the modulated phases.
This is valid at the transition but it is not excluded that higher harmonics
may appear at lower temperatures, as secondary order parameters.
$c_\mu(q)$ and $s_\mu(q)$ are the $x$ and $y$ components of three polarization
vectors
\BE
{\bf S}_\mu(q) = c_{\mu}(q){\bf \hat{x}} + s_{\mu}(q){\bf \hat{y}}.
\EE
These are complex quantities that may be expressed as
\BE
c_\mu = |c_\mu|e^{i \varphi_\mu^c}
\quad\text{and}\quad
s_\mu = |s_\mu|e^{i \varphi_\mu^s}.
\EE
This choice of variables allows for any elliptical polarization and relative
global phase for each column. To simplify notation, we replace $|c_\mu|$ by
$c_\mu$ and $|s_\mu|$ by $s_\mu$ and, to avoid any ambiguity, we make no use
of the complex $c_\mu$ and $s_\mu$ anymore. In real space,
\BEA
\L {\bf S}_{\mu k} \R
&&= \case1/2[ {\bf S}_{\mu}(q) e^{iqz_k}
+ {\bf S}_{\mu}^\ast(q) e^{-iqz_k} ] \NN\\
&&= c_\mu\cos(qz_k+\varphi_\mu^c){\bf \hat{x}}
+ s_\mu\cos(qz_k+\varphi_\mu^s){\bf \hat{y}},
\EEA
with $z_k=k+\frac12\delta_{\mu 3}$, so that $\varphi_\mu^{c,s}$ are the global
phases of the different columns $\mu$ at the $z=0$ level.

The function to minimize is then $F=F_T+F_J$, with
\BEA
\label{ft0}
F_T = T \sum_\mu \Bigl\{&& 2( c_\mu^2 + s_\mu^2 )
+\case1/4 \Bigl[ 6( c_\mu^4 + s_\mu^4 ) \NN\\
&&+4 [2+\cos 2(\varphi_\mu^s-\varphi_\mu^c)] c_\mu^2 s_\mu^2
\Bigr] \Bigr\}
\EEA
and
\BEA
\label{fj0}
F_J = -\sum_{\mu\nu} \Bigl[&&
J_{\mu\nu}^c c_\mu c_\nu \cos(\varphi_\nu^c-\varphi_\mu^c) \NN\\
&&+ J_{\mu\nu}^s s_\mu s_\nu \cos(\varphi_\nu^s-\varphi_\mu^s)
\Bigr],
\EEA
with couplings as previously defined. $F$ is a function of $c_\mu$, $s_\mu$,
$\varphi_\mu^{c,s}$ and $q$ that, at first, is numerically minimized. The
reader should note that the permutation $1\leftrightarrow 2$ in the indices
leaves $F$ unchanged, which reflects the equivalence of columns 1 and 2.

\subsubsection{Helicity patterns}
We numerically observe simple relationships between the phases $\varphi_\mu^c$
and $\varphi_\mu^s$. These, in turn, lead to a simplified expression for the
free energy functional. For $G=0$, the cosine and sine components
are equivalent and $s_\mu = c_\mu$. By numerically minimizing $F$, we obtain
\BE
\varphi_\mu^s-\varphi_\mu^c = \pm \case\pi/2,
\EE
so that the modulation appearing in the columns is circularly polarized, with
a helicity given by the sign on the right-hand side. This sign
($\sigma_\mu=\pm$) may differ from one $\mu$ value to another. We denote
$(\sigma_1,\sigma_2,\sigma_3)$ the helicity configuration of the three
sublattices of columns. For $G=0$, the only allowed helicity configuration is
$(+++)$ (or equivalently $(---)$). Depending on the sign of $J$, the relative
global phases of each component from one column to another are 0 or $\pm\pi$.
For $J>0$, the columns adopt a ``ferromagnetic-like'' arrangement:
\BE
\varphi_1^{c,s} = \varphi_2^{c,s} = \varphi_3^{c,s},
\EE
and for $J<0$, the triangular geometry imposes a colinear ``antiphase'':
\BEA
\varphi_1^{c,s} = \varphi_2^{c,s} = \varphi_3^{c,s}\pm\pi.
\EEA
Instead of the ordinary ``$120^\circ$ state'' for an evenly frustrated system,
our system concentrates the frustration in the 1-2 bond, which has a lower
coordination number than 1-3 and 2-3 bonds. In contrast to the results
obtained at $T=0$ in \cite{Hebert96}, where every column was forced to adopt
the same amplitude and a deformed ``$120^\circ$ state'' was achieved,
we obtained here a fully colinear antiphase.

\begin{figure}
\vglue 0.4cm\epsfxsize 8cm\centerline{\epsfbox{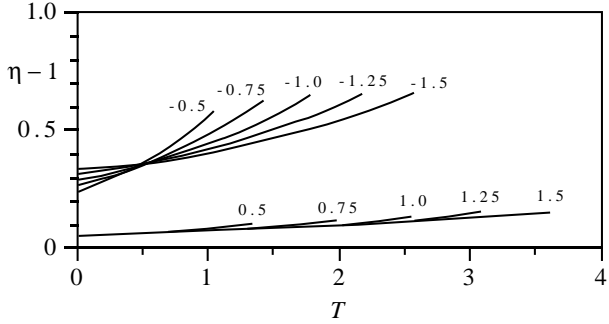}}\vglue 0.4cm
\caption{\baselineskip=12pt
$\eta(T)$ curves for $G=0$ and different values of $J$. $\eta(T)$ is defined only for $T$ below the critical temperature $T_c$, depending on the $J$ value.
}\label{eta}
\end{figure}

This fundamental difference arises from the freedom of the above model
to adopt different amplitudes of modulation for the different columns. This
was not allowed in \cite{Hebert96}. Using the following definitions:
\BE
\eta_c = {c_3 \over c_1} = {c_3 \over c_2}
\qquad\qquad
\eta_s = {s_3 \over s_1} = {s_3 \over s_2},
\EE
we have presented on Fig.~\ref{eta}, for $G=0$ and for different values of
$J$, $\eta_c$ and $\eta_s$ as functions of the temperature. Because $G=0$,
$\eta_c=\eta_s=\eta$ and the curves are identical for each $J$. This ratio
has a well-defined value only below the critical temperature $T_c$. It is to
be noticed that for all cases presented, $\eta$ is always larger than unity.
For $J<0$, the behavior is even larger than for $J>0$. As a consequence, the
displaced columns show a larger amplitude for the modulated phases.

\begin{figure}
\vglue 0.4cm\epsfxsize 5cm\centerline{\epsfbox{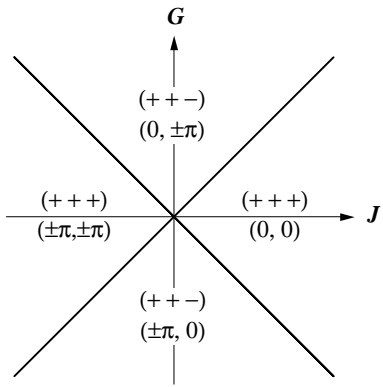}}\vglue 0.4cm
\caption{\baselineskip=12pt
Helicity patterns and phase differences in the $J$-$G$ plane. The first number (0 or $\pm\pi$) is the difference $\varphi_3^c - \varphi_1^c$, and the second is $\varphi_3^s - \varphi_1^s$.
}\label{helicite}
\end{figure}

For $G\neq 0$, the rotational invariance is broken and we expect
non-circularly polarized phases. Numerically, we still found
$\varphi_\mu^s-\varphi_\mu^c = \pm \case\pi/2$ and, if $|J|\ge|G|$, a $(+++)$
configuration is realized. If $|J|<|G|$, we have a $(++-)$ configuration, as
previously found in \onlinecite{Plumer93} (see Fig.~\ref{helicite}). The
inter-column relative phases are related to the sign of $J+G$ for the cosines
and of $J-G$ for the sines. For $J+G>0$,
\BE
\label{conf1}
\varphi_1^c = \varphi_2^c = \varphi_3^c,
\EE
while for $J+G<0$,
\BE
\label{conf2}
\varphi_1^c = \varphi_2^c = \varphi_3^c\pm\pi.
\EE
For $J-G>0$,
\BE
\label{conf3}
\varphi_1^s = \varphi_2^s = \varphi_3^s,
\EE
while for $J-G<0$,
\BE
\label{conf4}
\varphi_1^s = \varphi_2^s = \varphi_3^s\pm\pi.
\EE

These phase relationships divide the $J$-$G$ plane in four quadrants, as seen
in Fig.~\ref{helicite}. This diagram possesses an athermal character since
these relative phases are the only variables having an influence on the sign
of each term of $F$, $\cos\case1/2 q$ being always positive. However, it is
important to stress that the phases (and the phase relationships) have a
physical meaning only when their corresponding amplitudes are nonzero. From
these partial results, we may rewrite $F$ in a simpler form:
\BEA
\label{ft}
F_T =&& 2T\{ 2c_1^2 + c_3^2 + 2s_1^2 + s_3^2 \NN\\
&&+ \case1/4[ 3( 2c_1^4 + c_3^4 + 2s_1^4 + s_3^4 )
+ 2( 2c_1^2 s_1^2 + c_3^2 s_3^2 ) ] \},
\EEA
\BEA
\label{fj}
F_J = &&-(\cos q + J_2\cos 2q)( 2c_1^2+c_3^2+2s_1^2+s_3^2 ) \NN\\
&&-3(J+G)c_1^2 -12|J+G|(\cos\case1/2 q)c_1 c_3 \\
&&-3(J-G)s_1^2 -12|J-G|(\cos\case1/2 q)s_1 s_3. \NN
\EEA
The absolute values of some couplings reflects the helicity choice made in
order to minimize $F$. The positive signs of the coefficients of $c_1^2 s_1^2$
and $c_3^2 s_3^2$ favorize competition between $c$ and $s$ variables.

\subsubsection{Phase boundaries at $T_c$}
In this section, we investigate the first order-disorder transition to appear
when lowering the temperature, that is, the highest temperature for which at
least one order parameter is nonzero. This resulting critical temperature and
the boundary between different phases should agree with the results found in
Section A. This is presented in order to express the consistency of the Landau theory
and its numerical treatment. For each parameter set $(T,J,G,J_2)$, we minimize
$F$ defined with the expressions (\ref{ft}) and (\ref{fj}); using (\ref{ft0})
and (\ref{fj0}) would be equivalent, but numerically inefficient. The critical
temperature is found by selecting $T$ such that one of $c_1$, $c_3$, $s_1$ or
$s_3$ is as small as possible, but nonzero. We have recovered the diagram of
Fig.~\ref{diagTc}: $T_c$ and $q_c$ being identical to those found in Section
A. We also conclude that the transitions are always of second order, even with
sixth-order terms in the expansion of $F_T$.

For $q_c=0$, that is, when $|J|$ or $|G|$ are large enough, some analytical
results as easily obtainable. The matrix $\{H_F\}_{ij}=\{\partial^2 F/\partial
x_i\partial x_j\}$, where ${\bf x}$ is a vector constructed from the order
parameters (${\bf x} = (c_1,c_3,s_1,s_3)$), defines the local convexity of
$F$. We diagonalize $H_F$ to express this convexity along some
eigendirections: $h_F$, the four eigenvalues of $H_F$, are the convexity
coefficients. For each parameter set $(T,J,G,J_2)$, $F$ is minimal for a
particuliar ${\bf x}$. For the algebric form of the present $F$, the sign of
each convexity coefficient at ${\bf x}={\bf 0}$ indicates whether ${\bf 0}$ is
a minimum or not: $F$ is minimal at ${\bf 0}$ if every coefficient is
positive, but one or more negative coefficients means that ${\bf 0}$ is no
longer a minimum. A vanishing eigenvalue of $H_F$ at ${\bf x}={\bf 0}$
corresponds to a second-order transition.

$H_F$ has two distinct, doubly degenerate eigenvalues given by
\BEA
h_F^{c,s} =&& 3[2T-(1+J_2)-(J\pm G)] \NN\\
&& +\sqrt{[2T-(1+J_2)-3(J\pm G)]^2 + 144(J\pm G)^2}.
\EEA
$h_F^c$ correspond to $J+G$ and $h_F^s$ to $J-G$. For a set $(J,G,J_2)$, $T$
is the critical temperature $T_c$ when every $h_F$ is positive at ${\bf
x}={\bf 0}$, but at least one vanishes. The eigenvalues vanish at the
following temperatures:
\BE
T_c^{c,s} = \case1/2(1+J_2)
+\case3/8(J\pm G)+\case3/8|J\pm G|\sqrt{33}.
\EE
We kept the solution that keeps $T_c$ positive for any $J$ and
$G$. These expressions coincide with (\ref{relationTc}) if $q_c=0$. The cosine
or sine components order depending on which of the two temperatures $T_c^c$ or
$T_c^s$ is maximum. We have a phase boundary between the two types of order
for $T_c^c=T_c^s$. This equality is equivalent to (\ref{relationJG}). For
$q_c\ne 0$, the same analysis leads easily to (\ref{relationTc}).

For a small $q_c$, we expand $T_c^{c,s}$ in powers of $q_c^2$. The $q_c^4$
coefficient is negative. The boundary between $q_c=0$ and $q_c\ne 0$ phases
occurs when the $q_c^2$ coefficient vanishes. We have the relation
\BE
|J\pm G| = {22\over\sqrt{33}}(-J_2-\case1/4),
\EE
which defines the boundaries on Fig.~\ref{diagTc} (the dashed lines).

\subsubsection{Thermal phase diagrams}

Below the critical temperature, and for $J$ and $G$ exactly lying on the
$|G|=-g(J,J_2)J$ curve, both the $c$ and $s$ components order. For the same
$G$ and $J_2$ but for a higher $J$, the $c$ components order first, followed
at lower temperature by the $s$ components, and conversely for a lower $J$.
Consequently, the concomitant ordering of both the $c$ and $s$ components
exists for a range of $J$. From the equivalence of the $\mu=1$ and $\mu=2$
columns, we have observed that $c_1=c_2$ and $s_1=s_2$. Conversely, but
because of their mutual hindrance, $c_1<c_3$ and $s_1<s_3$.

\begin{figure}
\vglue 0.4cm\epsfxsize 7cm\centerline{\epsfbox{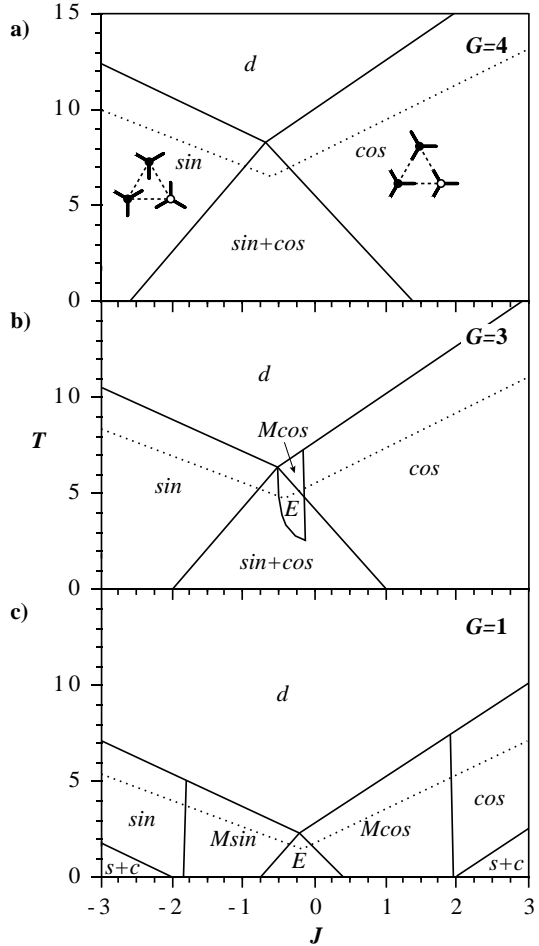}}\vglue 0.4cm
\caption{\baselineskip=12pt Thermal phase diagrams for strong transverse
couplings (see text). For $G=4$ (a) there is no modulated phases. For $G=3$
(b) the modulated phases are reentrant. The dotted lines indicate the limit of
validity of the free energy limited power series expansion, below which at
least one order parameter is greater than 0.5. The insets schematically
illustrate the orientations of the three molecules in a plaquette for {\it
sin} and {\it cos} phases. }\label{diagFort}
\end{figure}

We show in Fig.~\ref{diagFort} the thermal phase diagrams for constant values
of $G$ and $J_2=-1$. The symmetry of $F$ with respect to $G\leftrightarrow -G$
allows us to consider only $G\ge 0$. Each diagram has a high-temperature
disordered phase (denoted {\it d}) and, at lower temperatures, ordered phases
{\it sin} and {\it cos}. For $q\ne 0$, these phases are denoted {\it Msin} and
{\it Mcos}. Just below the $|G|=-g(J,J_2)J$ curve is a mixed {\it sin}+{\it
cos} phase that transforms continuously from {\it sin} to {\it cos} when
approaching this region from large positive and negative $J$. For $q\ne 0$,
the region {\it sin}+{\it cos} is an elliptical phase denoted $E$.

For high values of $|J|$ or $|G|$, $q$ remains zero below $T_c$, as shown in
Fig.~\ref{diagFort}a. For small $|G|$ and increasing $|J|$, steep boundaries
between modulated and non-modulated phases are crossed (see
Fig.~\ref{diagFort}c): $q$ vanishes continuously from {\it Mcos} to {\it cos}
and from {\it Msin} to {\it sin}. Inversely, for small $|J|$ and increasing
$|G|$, a reentrant boundary is crossed: a modulated elliptical phase reappears
with increasing temperatures (see Fig.~\ref{diagFort}b). For high values of
$|J|$, below the critical temperature, the disordered variables ($c$ in the
{\it sin} phase and $s$ in the {\it cos} phase) order at lower temperatures,
forming {\it sin}+{\it cos} phases. For $G=1$ (Fig.~\ref{diagFort}c), we see
the beginning of this phase for $|J|>\sim 2$. The phase boundaries are
parallel to the {\it sin} and {\it cos} boundaries with the disordered phase.
These low temperature phases are not shown on Figs \ref{diagFort}a and
\ref{diagFort}b, but they occur respectively for $|J|>\sim 6$ and 8. Because
helicity configurations have meaning only when $c_\mu$, $s_\mu$ and $q$ are
nonzero, that is, for $E$ phases, the diagrams of Fig.~\ref{diagFort} do not
show the richness of Fig.~\ref{helicite}. For the choice of parameters, the
central $E$ phases in Fig.~\ref{diagFort} are $(++-)$ phases. For smaller
values of $|G|$, it is expected that a phase boundary between $E$ phases
having $(++-)$ and $(+++)$ configurations would be observable.

\begin{figure}
\vglue 0.4cm\epsfxsize 7cm\centerline{\epsfbox{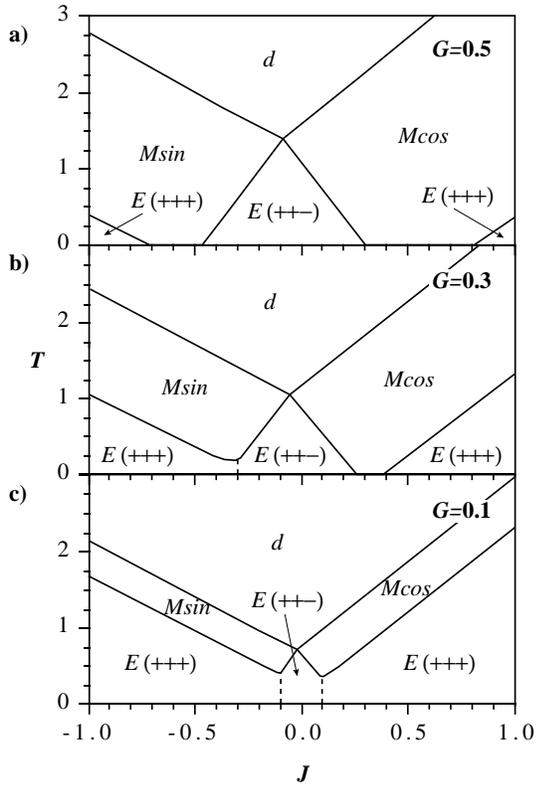}}\vglue 0.4cm
\caption{\baselineskip=12pt Thermal phase diagrams in the
quasi-onedimen\-sional regime (see text). For $G=0.5$ (a) the $E(++-)$ and
$E(+++)$ are separated. For $G=0.3$ (b) and $G=0.1$ (c), dashed lines denote
first-order helicity reversal transitions. }\label{diagFaible}
\end{figure}

Indeed, the quasi one-dimensional regime, where $|J_1|,|J_2|\gg |J|,|G|$, is
interesting because it is more realistic for HHTT, with $E$ phases having
different helicity patterns. We show in Fig.~\ref{diagFaible} thermal phase
diagrams for $G$ from 0.1 to 0.5. Both the $E(+++)$ and $E(++-)$ phases exist
and are sometimes adjacent. On the $J$ interval presented in
Fig.~\ref{diagFaible}, all the ordered phases are modulated. The temperature
``depth'' of the {\it Msin} and {\it Mcos} phases is approximately
proportional to $|G|$ and independent of $J$.

\begin{figure}
\vglue 0.4cm\epsfxsize 8cm\centerline{\epsfbox{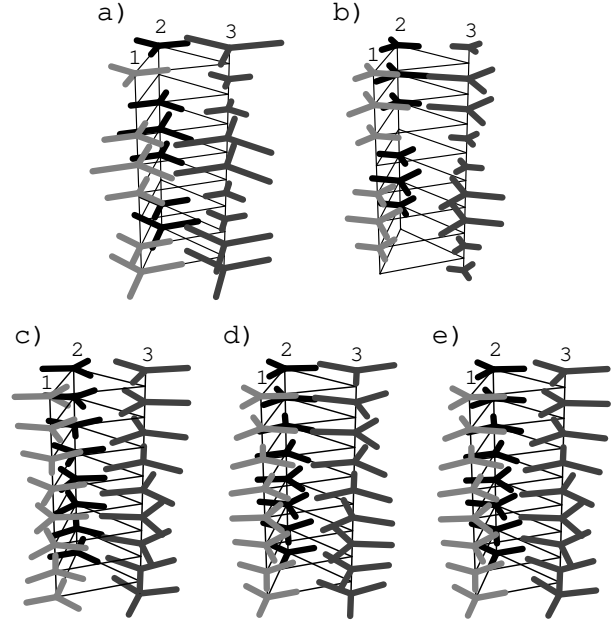}}\vglue 0.4cm
\caption{\baselineskip=12pt Schematic illustration of the orientations of
molecules and amplitudes of mean values with $G>0$ for a) the {\it Msin} phase
with column 3 in phase opposition, b) the {\it Mcos} phase with column 3 in
phase conjunction, c) the $E(+++)(\pi,\pi)$ phase, d) the $E(++-)(0,\pi)$
phase and e) the $E(+++)(0,0)$ phase. }\label{illustration}
\end{figure}

Fig.~\ref{illustration} illustrates schematically the positions and
orientations of the molecules for different phases encountered in the diagrams
of Fig.~\ref{diagFaible}, in the case where $G>0$. Some are also present in
Fig.~\ref{diagFort}. The length of the tails represent the amplitude of
ordering. The {\it Msin} phase (Fig.~\ref{illustration}a) shows modulated
order with uniform orientations along the columns and where column 3 is in
phase opposition. For $G>0$, the {\it Mcos} phase (Fig.~\ref{illustration}b)
is a modulated ferromagnetic-like state. The maximum lengths of the tails
shows that column 3 is more ordered than columns 1 and 2. The corresponding
{\it sin} and {\it cos} phases are illustrated in Fig.~\ref{diagFort}a.
Fig.~\ref{illustration}c to d represent the various elliptical phases $E$
encountered from left to right in Fig.~\ref{diagFaible}.

\section{Monte Carlo simulations}
%
\subsection{Method and algorithm}

We use the Metropolis algorithm (see \onlinecite{Binder79} on general
Monte Carlo methods) to simulate a three-dimensional lattice of XY variables.
Our goal is to obtain the essential features of the thermal phase diagram of
the system (\ref{hamilCos}), restricting ourselves to the
quasi one-dimensional regime.

The above method has been implemented in conjunction with the ``spiraling''
algorithm\cite{Saslow92,Collins96}, which simulates incommensurate
helical phases by relaxing the constraint associated with the finite size
of the lattice. Outside the critical regime, the results are in principle
independent of the lattice size. This method is applied on each column of $N$
XY variables $\{\phi_1, \dots, \phi_N\}$. We simulate the neighbors of
$\phi_1$ and $\phi_2$ with {\it phantom} sites (denoted by a prime) related to
sites of the opposite extremity of the column:
\BE
\phi'_{N-1} = \phi_{N-1} - N\Delta,
\qquad
\phi'_N = \phi_N - N\Delta.
\EE For the sites $\phi_{N-1}$ and $\phi_N$, we use the phantom neighbors
\BE
\phi'_1 = \phi_1 + N\Delta,
\qquad
\phi'_2 = \phi_2 + N\Delta.
\EE The reader will note that, for $\Delta=0$, simple periodic boundary
conditions are recovered. $\Delta$ is an effective field representing an
additional indefinite length of a lattice modulated with a constant pitch. For
the system to select its own boundary conditions, we consider $\Delta$ as a
thermodynamic variable. Each Monte Carlo step of the spiraling algorithm
consists in $N$ ordinary trial flips of the $\phi$ variables and a single trial
flip of $\Delta$. We modify $\Delta$ by a small random angle $\delta\Delta$.
In order for the new $\Delta$ to be compatible with the above equations,
a twist must be imposed on the lattice:
\BE
\phi_i \leftarrow \phi_i + (i-1)\delta\Delta.
\EE The new state has a new total energy, and is then tested for acceptance
with the Metropolis algorithm.
We may compare the spiraling algorithm to a high-order mean-field
approximation, where the exact clusters have the size of the finite lattice
used in the simulation. Close to a transition, when the correlation length is
large relative to the cluster size, this approximation loses its validity

We simulate a $6\times 6$ triangular lattice of columns, that is, 12 groups of
three columns. We use periodic boundary conditions in the plane and the
spiraling algorithm in the columns' direction. Each column has its own
$\Delta$ variable.

For two neighboring columns with non-planar interactions ($J'$ and $G'$), the
spiraling algorithm introduces some small energy discrepancies. This problem
arises from the non-equivalence of some of the couplings at the edges of the
lattice. Due to the relatively small energy involved in this boundary effect
(especially for a quasi one-dimensional system), it was neglected.

We identify the different phases with the nonzero values of the Fourier
coefficients of the $x$ and $y$ components defined by:
\BML
\BE a_\mu^c(\tilde{q}) = {2\over 12N} \sum_{(i,j)\in R_\mu} \sum_k
\cos\phi_{ijk} \cos\tilde{q}z_k,
\EE
\BE b_\mu^c(\tilde{q}) = {2\over 12N} \sum_{(i,j)\in R_\mu} \sum_k
\cos\phi_{ijk} \sin\tilde{q}z_k,
\EE
\BE a_\mu^s(\tilde{q}) = {2\over 12N} \sum_{(i,j)\in R_\mu} \sum_k
\sin\phi_{ijk} \cos\tilde{q}z_k,
\EE
\BE b_\mu^s(\tilde{q}) = {2\over 12N} \sum_{(i,j)\in R_\mu} \sum_k
\sin\phi_{ijk} \sin\tilde{q}z_k,
\EE
\EML with $z_k=k+\frac12\delta_{\mu 3}$. $R_\mu$ represents the sublattice of
$\mu$ columns and $N$ is the number of sites in each column. Using these
coefficients, we construct the order parameters:
\BML
\label{amplitudes}
\BE c_\mu(\tilde{q}) = \sqrt{a_\mu^c(\tilde{q})^2 + b_\mu^c(\tilde{q})^2},
\EE
\BE s_\mu(\tilde{q}) = \sqrt{a_\mu^s(\tilde{q})^2 + b_\mu^s(\tilde{q})^2}.
\EE
\EML The $\L c_\mu(\tilde{q})\R$ and $\L s_\mu(\tilde{q})\R$ profiles show a
maximum at $\tilde{q}=q_0$, and may show secondary peaks. Indeed, it has been
shown\cite{Walker80} that a third harmonic at $q_3=3q_0$ should appear for
linearly polarized columns. In our simulations, this third harmonic should be
an indication of such a linear polarization. We also construct the relative
angular phases of the columns defined by:
\BML
\label{phases}
\BE
\cos(\phi_\mu^s-\phi_\mu^c)(\tilde{q}) = {a_\mu^s(\tilde{q})
a_\mu^c(\tilde{q}) + b_\mu^s(\tilde{q}) b_\mu^c(\tilde{q})
\over s_\mu(\tilde{q}) c_\mu(\tilde{q})},
\EE
\BE
\sin(\phi_\mu^s-\phi_\mu^c)(\tilde{q}) = {b_\mu^s(\tilde{q})
a_\mu^c(\tilde{q}) - a_\mu^s(\tilde{q}) b_\mu^c(\tilde{q})
\over s_\mu(\tilde{q}) c_\mu(\tilde{q})},
\EE
\BE
\cos(\phi_{\mu+1}^c-\phi_\mu^c)(\tilde{q}) = {a_{\mu+1}^c(\tilde{q})
a_\mu^c(\tilde{q}) + b_{\mu+1}^c(\tilde{q}) b_\mu^c(\tilde{q})
\over c_{\mu+1}(\tilde{q}) c_\mu(\tilde{q})},
\EE
\BE
\sin(\phi_{\mu+1}^c-\phi_\mu^c)(\tilde{q}) = {b_{\mu+1}^c(\tilde{q})
a_\mu^c(\tilde{q}) - a_{\mu+1}^c(\tilde{q}) b_\mu^c(\tilde{q})
\over c_{\mu+1}(\tilde{q}) c_\mu(\tilde{q})},
\EE
\BE
\cos(\phi_{\mu+1}^s-\phi_\mu^s)(\tilde{q}) = {a_{\mu+1}^s(\tilde{q})
a_\mu^s(\tilde{q}) + b_{\mu+1}^s(\tilde{q}) b_\mu^s(\tilde{q})
\over s_{\mu+1}(\tilde{q}) s_\mu(\tilde{q})},
\EE
\BE
\sin(\phi_{\mu+1}^s-\phi_\mu^s)(\tilde{q}) = {b_{\mu+1}^s(\tilde{q})
a_\mu^s(\tilde{q}) - a_{\mu+1}^s(\tilde{q}) b_\mu^s(\tilde{q})
\over s_{\mu+1}(\tilde{q}) s_\mu(\tilde{q})}.
\EE
\EML We compare $\L\cos(\phi_\mu^s-\phi_\mu^c)(q_0)\R$ to the mean-field
expression $\cos(\phi_\mu^s-\phi_\mu^c)$, and so on for each mean value.

The amplitudes (\ref{amplitudes}) converge relatively fast, but the angular
phases (\ref{phases}) converge much more slowly: they are ratios of
fluctuating quantities. For some simulations, we are only interested in the
$\L c_\mu(\tilde{q})\R$ and $\L s_\mu(\tilde{q})\R$ values, and we have
simulated only 2000~MCS/S, including 1000~MCS/S for thermalization. $N$, the
number of sites per column, was taken to be 40. However, many more steps was
needed to obtain the relative angular phases (25000~MCS/S, including
5000~MCS/S to thermalize). In these cases, $N$ was taken to be 12. To improve
numerical efficiency, we discretize the $\phi$ and $\Delta$ variables into 256
values from 0 to $2\pi$.

All simulations are done with $G=0.1$ and $J_2=-1$, a quasi one-dimensional
limit. We perform the simulations by gradually decreasing the temperature for
each value of $J$. This allows a greater numerical stability for
low-temperature phases and retains the helicity sign ($(++\mp)$ or $(--\pm)$)
for all temperatures.

Preliminary simulations on a single plaquette of three columns were done to
compare the spiraling algorithm with periodic boundaries conditions in some
reasonable computing time. For $N=40$, simulations using periodic boundaries
along columns give very similar results compared to simulations with the
spiraling algorithm, except for smaller mean values (because of higher
fluctuations) and slightly displaced peaks. However, the difference is
pronounced for $N=12$, where the spiraling algorithm broadens the principal
peak, while periodic boundaries conditions destroy the whole spectrum.

\begin{figure}
\vglue 0.4cm\epsfxsize 8cm\centerline{\epsfbox{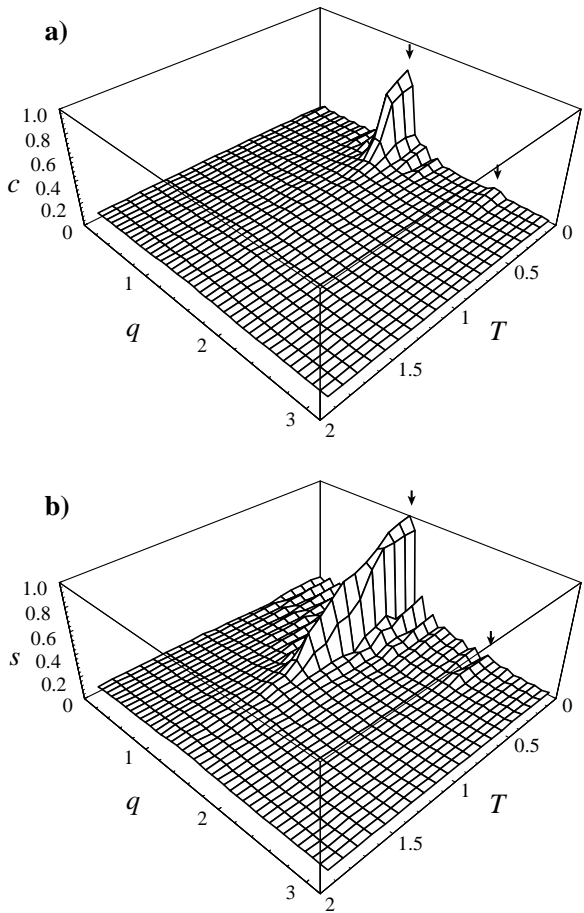}}\vglue 0.4cm
\caption{\baselineskip=12pt a) $\L c_1(q)\R(T)$ and b) $\L s_1(q)\R(T)$
profiles for $J=-0.15$, $G=0.1$ and $J_2=-1$. The arrows indicate the first
and third harmonics. The $q_0$ peaks appear at $T\approx 0.4$ for $c_1$ and
$T\approx 0.9$ for $s_1$. }\label{profils}
\end{figure}

\subsection{Thermal phase diagrams}

Fig.~\ref{profils} shows a typical result for the order parameters, from
which the thermal phase diagram is constructed: the profiles $\L
c_1(\tilde{q})\R(T)$ and $\L s_1(\tilde{q})\R(T)$. Notice that both $q_0$ and
$q_3$ peaks appear at specific temperatures ($q_3$ is folded in the $[0,\pi]$
interval). The $\L c_\mu(\tilde{q})\R$ and $\L s_\mu(\tilde{q})\R$ profiles
show that, for $J=-0.15$, a decreasing temperature drives the system from {\it
d} phase, to {\it Msin} phase and finally to $E$ phase. The critical
temperatures are arbitrarily taken to be the points at which the amplitude of
the peak is half its maximum value.

\begin{figure}
\vglue 0.4cm\epsfxsize 8cm\centerline{\epsfbox{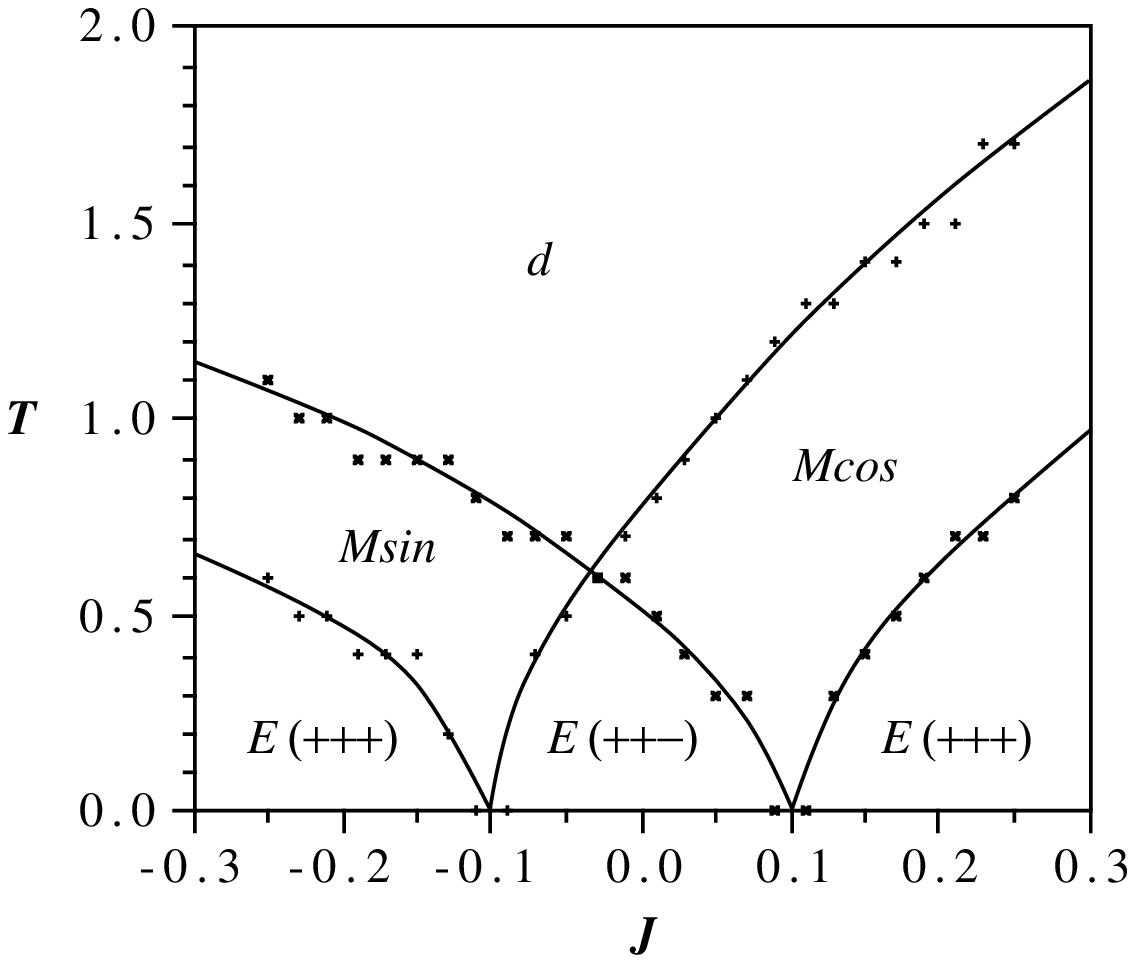}}\vglue 0.4cm
\caption{\baselineskip=12pt Thermal phase diagram for $G=0.1$ and $J_2=-1$
from Monte Carlo simulations. The continuous line is a guide to the eye (see
Fig.~\ref{diagFaible}c for the mean-field diagram). }\label{diagMC}
\end{figure}

The thermal phase diagram is constructed by repeating the simulations for many
$J$ values. Fig.~\ref{diagMC} shows the diagram for $G=0.1$ and $J_2=-1$. We
always obtain $\L\cos(\phi_\mu^s-\phi_\mu^c)(q_0)\R \approx 0$ and
$\L\sin(\phi_\mu^s-\phi_\mu^c)(q_0)\R \approx \pm 1$, with signs corresponding
to the mean-field helicity configurations: $(++-)$ or $(--+)$ for $|J|<0.1$
and $(+++)$ or $(---)$ for $|J|>0.1$. Some longer simulations were done to
establish clearly the phase differences, at $J=-0.15$, $-0.05$, $0.05$ and
$0.15$. Given some high-enough $\L c_\mu(q_0)\R$ and $\L s_\mu(q_0)\R$
amplitudes, every relative phase is compatible with the mean-field
calculations and the results just cited. Moreover, many short simulations
where done on a single plaquette of three columns ($N=40$) and, although no
phase transition is observed (the system is onedimensional), the phase
differences are in excellent agreement with Fig.~\ref{helicite}.

Contrarily to the mean-field results (Fig.~\ref{diagFaible}c), the $E$ phases
with different helicity configurations were not found to be adjacent: a
linearly polarized phase opens up between the $E(+++)$ and $E(++-)$ phases. On
the diagram, for a given value of $J$, a dot at $T=0$ means that no phase
transition was clearly identified when lowering the temperature.

\section{Discussion and conclusion}

X-ray measurements on HHTT have lead to the observation of a sequence of
phases as temperature is lowered. Two neighboring phases ($D_{hd}$ and $H$) at
intermediate temperatures have a triangular structure of columns of molecules.
Entering the $H$ phase from the $D_{hd}$ phase involves a mutual and
concomitant ordering of both the position and orientation of the molecules
along the columns. However, it is immediately realized that the positions of
the molecules are rapidly frozen as compared to their orientations, being
submitted to more stringent intermolecular forces. This is the main
justification for the model studied in this paper, limited to orientational
degrees of freedom. The obtained results  and in particular the general trend
of the thermal phase diagram should be useful to understand the behavior of
HHTT inside the $H$ phase.

The first result of interest is the ordering of linearly polarized phases at
$T_c$ with a finite wavenumber for the amplitude modulation in the columnar
direction, whose value is decreasing with increasing values of the
intermolecular couplings and eventually vanishes. A similar disappearance
of the amplitude modulation has been predicted\cite{Hebert96} at $T=0$ where
it then shows up as an unwinding of the helical pitch. At $T_c$, for large
$|J|$ and $|G|$, a boundary at a constant slope of magnitude $\sqrt{33}$ is
predicted between two linearly polarized phases. It is to be noticed that this
behavior may be seen as a reminiscent effect of the two boundary structures
predicted at $T=0$ between linearly polarized phases. At $T=0$, the constant
slopes of the boundaries are respectively $-3$ and $-5$. As seen on
Fig.~\ref{diagFort}a, this $T=0$ behavior is perfectly predictible at large
$G$.

The frustration between ordered columns of molecules with an octupolar
moment on a triangular lattice is very high. Part of this frustration is
relaxed by freezing the positions of the molecules and displacing one of every
three columns by half a lattice spacing in the columnar direction. However,
for negative value of the interaction parameter $J$, even under these
conditions, substantial orientational frustration remains. At $T=0$, with
only constant amplitude phases, a non-colinear distorted $120^\circ$ phase was
obtained with global relative angular phases between the columns which
depended on the pitch of the helical modulation of the columns. In the present
case, allowing for different amplitudes of the modulation for the displaced
columns compared to the undisplaced columns, the resulting configuration is
colinear and the relative angular phases is independent of the pitch. This
angular configuration, typical of unfrustrated systems, is only achieved
through a larger amplitude of modulation for the displaced columns as compared
to the undisplaced ones, as shown if Fig.~\ref{eta}.

For moderate values of $G$ as compared to the intracolumnar couplings
(Fig.~\ref{diagFaible}), when there exists elliptical phases $E$, both the
helicity patterns $(+++)$ and $(++-)$ are predicted. Also it is to be noted
that the temperature range over which the linearly polarized and modulated
phases exist, narrows down with decreasing value of $G$. Under these
conditions, it is expected that we rapidly enter the elliptical phases on
lowering the temperature below $T_c$. In this temperature range, where our
model, which freezes the positional degrees of freedom rapidly under $T_c$, is
certainly valid, and for $|G|>|J|$, the elliptical phase $(++-)$ is predicted.
Recall that the $(++-)$ phase is the one observed\cite{Fontes88} for HHTT in
the $H$ phase. We then conclude that HHTT is a quasi one-dimensional system
where the molecular orientations are dominated by the non-rotationally
invariant interaction between octupolar moments located on a distorted
triangular lattice.

The finite temperature Monte Carlo simulations on finite size systems has
confirmed remarkably the mean field results. The one important difference is
that the $(++-)$ and $(+++)$ configurations are not seen to be adjacent in
Monte Carlo simulations where they were predicted to have a common boundary in
the mean field approximation. The hard-spin constraint $|{\bf S}_{\mu m}|=1$
is automatically satisfied in the Monte Carlo simulations. At low temperature
and in the mixed regime $|J|\approx|G|$, the system is indecisive between
configurations where all the columns have identical helicities and opposite
helicities between the displaced columns and the undisplaced columns. Based on
the hard-spin contributions which are optimal for non-helical phases, the
system seems to be driven to zero temperature while retaining either a
modulated  {\it cos} or {\it sin} phase according to the sign of $J$. This
result probably manifest itself into the rare occurence of helical phases
experimentally.

Regarding the $D_{hd}\rightarrow H$ transition in HHTT, this work raises the
question of the detailed nature of the observed $H$ phase near the transition:
is there a linearly polarized ``$H$ phase''? The presence of a linearly
polarized modulated phase would be an indication of the chiral octupolar
nature of the molecule (the $G$ coupling). Such phases could be observed as a
third harmonic in the modulation of the columns in X-ray measurements. The
most important result of this paper is the higher amplitude predicted for the
displaced columns as compared to the undisplaced columns. This predicted
behavior would affect the X-rays results and in particular the relative
amplitude and thermal broadening of the Bragg's peaks. The $J$ and
$G$ values may be varied by using discoid compounds other than HHTT or by the
application of surface pression.

\acknowledgements
The authors are grateful to M.L. Plumer for many discussions. G.L. would like
to thank W.M. Saslow for helpful suggestions related to the simulations. This
work is supported by NSERC (Canada) and by FCAR (Qu\'ebec).

%

\end{document}